\documentclass[a4paper,twoside]{article}
\newcommand{\affil}[1]{$^{\rm #1}$}
\usepackage[authoryear]{natbib}
\bibpunct{(}{)}{;}{a}{}{,}
\usepackage{graphicx}
\date{} 
\title{\large\bf\flushleft The Nature of The Light Variations of Unique Binary DK CVn}
\author{\parbox{\textwidth}{\flushleft
\vspace{-0.5cm}
{\it Hasan Ali DAL \affil{1},\affil{2}, Esin S\.{I}PAH\.{I} \affil{1}, Orkun \"{O}ZDARCAN} \affil{1} \\
\vspace{0.4cm}
{\small \affil{1}\,Department of Astronomy and Space Sciences, University of Ege, Bornova, 35100 ~\.{I}zmir, Turkey}\\
{\small \affil{2}\,Corresponding Author, Email: ali.dal@ege.edu.tr}}}
\begin{document}
\twocolumn[
\begin{changemargin}{.8cm}{.5cm}
\begin{minipage}{.9\textwidth}
\vspace{-1cm}
\maketitle
%
%
\small{\bf Abstract:}  In this study, we present the BVR observations of DK CVn in 2007 and 2008. We analysed the BVR light curves of the system and obtained the system's parameters. Using the "q-search" method, we estimated the mass ratio of the system ($q$) as 0.55. Taking the temperature of the primary component as 4040 K, the temperature of the secondary was found to be 3123 K in the analyses. Several flares were detected in this study, and the distributions of flare equivalent duration versus flare total duration are modelled using the One Phase Exponential Association function for these flares. The parameters of the model demonstrated that the flares are the same with analogous detected from UV Ceti type stars. Apart from flare activity, we reached some results about the sinusoidal-like variation at out-of-eclipse. The clues demonstrate that the variation at out-of-eclipse must be caused by some cool spot(s) on one of the components. The star is found to show two active longitudes in which the spots are mainly formed. Consequently, this study reveals that DK CVn should be a chromospherically active binary star.\\


\medskip
\medskip
\end{minipage}
\end{changemargin}
]
\small

\section{Introduction}

DK Canum Venaticorum (= GSC 03018 01509) is classified as an eclipsing binary star in the SIMBAD database. The system was discovered by the Robotic Optical Transient Search Experiment for the first time \citep{Ake00}.

Examining the light variation of the system, \citet{Die01} noted that the reflection effect is dominant in the light variation. DK CVn identified as a variable star by \citet{Kaz03} was listed as an Algol type variable (EA) in the Catalogue of Eclipsing Variables by \citet{Mal06}. \citet{Ter05} noticed from $UBVR_{C}I_{C}$ bands observations of the season 2002 that the amplitudes of the humps around $0^{P}.25$ increase from $I_{C}$ band to U band. A flare was detected by \citet{Ter05} in the observations of the season 2003. Using the low resolution spectroscopy, they determined that the primary component is from K7 spectral type. Considering the light curves obtained in $I_{C}$ band, they indicate that the secondary component must be a star from the late M spectral types.

In this study, BVR photometry of DK CVn was made in the seasons 2007 and 2008. All the details of the observations and reduction procedures are given in Section 2.1. The new light elements are given in Section 2.2. Using the PHOEBE V.0.31a software \citep{Prs05}, the BVR light curves were analysed, and some physical parameters were obtained. The details of the analysis are described in Section 2.3 and 2.4. Extracting the obtained synthetic light curve from the observed light curves, it was examined whether there is any variation out-of-eclipses. All the procedures are described in Section 2.5. Several flares were detected. Some parameters were computed for each flare. The One Phase Exponential Association (hearafter OPEA) model was derived for the flares. The analyses and the OPEA model of the flares are described in Section 2.6. All the results are given and discussed in Section 3.

\section{Observations and Analyses}

\subsection{Observations}

Observations were acquired with a thermoelectrically cooled ALTA U$+$42 2048$\times$2048 pixel CCD camera attached to a 40 cm - Schmidt - Cassegrains - type MEADE telescope at Ege University Observatory. The observations made in BVR bands were continued two nights in the season 2007 and five nights in the season 2008. Some basic parameters of program stars are listed in Table 1.

Although the program and comparison stars are very close on the sky, differential atmospheric extinction corrections were applied. The atmospheric extinction coefficients were obtained from observations of the comparison stars on each night. Moreover, the comparison stars were observed with the standard stars in their vicinity and the reduced differential magnitudes, in the sense variable minus comparison, were transformed to the standard system using procedures outlined by \citet{Har62}. The standard stars are listed in catalogues of \citet{Lan83, Lan92}. Furthermore, the dereddened colours of the system were computed. Heliocentric corrections were also applied to the times of the observations.

The mean averages of the standard deviations are $0^{m}.021$, $0^{m}.013$, and $0^{m}.019$ for observations acquired in the BVR bands, respectively. To compute the standard deviations of observations, we used the standard deviations of the reduced differential magnitudes in the sense comparisons minus check stars for each night. There was no variation observed in the standard brightnesses of the comparison stars.

In Figure 1, the light and colour curves of DK CVn are shown. Comparing the light curves of the seasons 2007 and 2008, it is obviously seen that there are some differences between two light curves. The decreases in the levels of both maxima and minima of 2007 and 2008 light curves are seen, and the shapes of the light curves also changed from 2007 to 2008. The parts of the light curves between $0^{P}.70$ and $1^{P}.20$ are especially different from each other. On the other hand, there is no variation in the colour curves except the flare moments.  Moreover, there are no variations in the mean colour, while there are some variations in the mean brightness. The secondary minimum was not obtained in 2007 due to the absence of the observation in those phases. In Figure 2, there are two V band light curves, whose data were taken from the database of The Northern Sky Variability Survey (hereafter NSVS) \citep{Woz04}. As seen from Figure 2, the shape of DK CVn's light curve is rapidly varying, even in one observing season.

\subsection{Times of Minima and Orbital Period}

We could find 54 times of minima (included both primary and secondary) from the literature. The data cover an interval from 2001 to the current time and listed in Table 2. All the minimum times used in the analysis were obtained in photoelectric observations. Using the linear least-squares method, an update linear ephemeris is given by Equation (1):

\begin{center}
\begin{equation}
HJD~=~24~53422.9838(2)~+~0^{d}.494964(1)~\times~E
\end{equation}
\end{center}

The linear correction $(O-C)_{I}$ is shown in Figure 3. The phases in all the figures are calculated with these new light elements.

\subsection{Light Curve Analysis}

Although the secondary minimum was not obtained in 2007, whole the light curve was obtained in 2008. Because of this, we analysed the BVR light curves obtained in 2008. The analyses were carried out with using the PHOEBE V.0.31a software \citep{Prs05}. The method used in the PHOEBE V.0.31a software depends on the method used in the version 2003 of the Wilson-Devinney Code \citep{Wil71, Wil90}. The BVR light curves were analysed with the "detached system", "semi-detached system with the primary component filling its Roche-Lobe" and "semi-detached system with the secondary component filling its Roche-Lobe" modes. An acceptable result was only obtained in "detached system" mode, while no acceptable results were obtained in all the others modes. \citet{Ter05}, who examined the low resolution spectrum of the system, demonstrated that the primary component is a K7V star. Besides, we took JHK brightness of the system ($J=10^{m}.489$, $H=9^{m}.839$, $K=9^{m}.664$) from the NOMAD Catalogue \citep{Zac05}. Using these brightnesses, we derived dereddened colours as a $(J-H)_{\circ}$=0$^{m}$.561 and $(H-K)_{\circ}$=0$^{m}$.140 for the system. Using the calibrations given by \citet{Tok00}, we derived the temperature of the primary component as 4040 K depending on these dereddened colours. The derived temperature is in agreement with the spectral type given by \citet{Ter05}. In the analyses, the temperature of the primary component was fixed to 4040 K, and the temperature of the secondary was taken as a free parameter.

Considering the spectral types, the albedos ($A_{1}$ and $A_{2}$) and the gravity darkening coefficients ($g_{1}$ and $g_{2}$) of the components were adopted for the stars with the convective envelopes \citep{Luc67, Ruc69}. The non-linear limb-darkening coefficients ($x_{1}$ and $x_{2}$) of the components were taken from \citet{Van93}. In the analyses, their dimensionless potentials ($\Omega_{1}$ and $\Omega_{2}$), the fractional luminosity ($L_{1}$) of the primary component and the inclination ($i$) of the system were taken as the adjustable free parameters.

In order to find the best photometric mass ratio of the components, we used the "q-search" method with using a step of 0.05 due to the absence of any spectroscopic mass ratios. As seen from Figure 4, the minimum sum of weighted squared residuals ($\Sigma res^{2}$) is found for the mass ratio value of $q=0.55$. According to this result, we assume that a possible mass ratio of the system is $q=0.55$.

As it is clearly seen from Figure 1, there is the clear asymmetry in the light curves of both 2007 and 2008. Moreover, the shape of the asymmetric light curve changed from 2007 to 2008. In order to remove the asymmetry, we assumed that the primary component has two cool spots on its surface. The synthetic light curves obtained from the best light curve solution are seen in Figure 5, and the result parameters of the analysis are also listed in Table 3. The 3D model of Roche geometry is shown in Figure 6.

\subsection{Estimated Absolute Parameters}

Although there is not any radial velocity curve of the system, we tried to estimate the absolute parameters of the components. Considering its spectral type, we took the mass of the primary component from \citet{Tok00}, and the mass of the secondary component was calculated from the estimated mass ratio of the system. Using Kepler's third law, we calculated possible the semi-major axis ($a$), and then the mean radii of the components were calculated. The mass of the primary component was found to be 0.44 $M_{\odot}$, and it was found to be 0.24 $M_{\odot}$ for the secondary component. Considering estimated $a$ value, the radius of the primary component was computed as 0.58 $R_{\odot}$, while it was computed as 0.59 $R_{\odot}$. Using the estimated radii and the obtained temperatures of the components, the luminosity of the primary component was estimated to be 0.08 $L_{\odot}$, and it was found as 0.03 $L_{\odot}$ for the secondary component.

The absolute parameters seem to be an acceptable as an astrophysical. However, the radii of the both components are larger than the expected values in respect to the theoretical models. In Figure 7, we plotted the distribution of the radii versus the masses for some stars. In the figure, the filled circles represent the known active stars, which were taken from the catalogue of \citet{Ger99}. Some of these stars exhibit the spot activity, while some of them exhibit the flare activity. Some stars exhibit both spot and flare activities. In the figure, the asterisk represents the secondary component, while the open triangle represents the primary component. The line represents the ZAMS theoretical model developed by \citet{Sie00}.

\subsection{The Variations Out-of-Eclipses}

In order to be sure of the reasons of the variations out-of-eclipses, we investigated the pre-whitened light curves in R band. For this aim, using the physical parameters, we derived the synthetic light curve of the system for unspotted case for the R band. Then, the synthetic light curve of R band was extracted from both R light curves obtained in 2007 and 2008. In the second step, the R band light curves, which were obtained in 2002, 2003, 2004 and 2005, presented by \citet{Ter05} were scanned, and the observational data were obtained from these light curves in order to compare our data with the data existing in the literature. Then, the synthetic light curve derived for unspotted case was extracted from Terrell's light curves.

The pre-whitened light curves of the season 2002, 2003, 2004, 2005, 2007 and 2008 are shown in Panels a, b, c, d, e and f in Figure 8, respectively. If the pre-whitened light curves are carefully examined, three points will be seen in the general nature of the light curves. One of them is a sinusoidal-like variation. The sinusoidal-like variations are seen in all the pre-whitened light curves as a dominant feature. The second one is that the sinusoidal-like variations have an asymmetric shape. Moreover, the shapes of the sinusoidal-like variations (the minima and maxima phases, their amplitudes and etc.) are varying from a year to the next one. Finally, there are some sudden and short-duration flares in the this curves.

In the pre-whitened light curves, one asymmetric minimum is seen in the sinusoidal-like variations, generally. However, there are two minima in the pre-whitened light curve of the season 2003. Examining these sinusoidal-like variations, some parameters were computed. These parameters are listed in Table 4. In the table, we listed the observing seasons, $\theta_{min}$, the amplitude of the pre-whitened R band light curves and references, respectively. The variation of the amplitude of the pre-whitened R band light curves is shown in Figure 9. As seen from the figure, although the amplitude of the pre-whitened light curves are decreasing from 2002 to 2008 in the general view, a cyclic behaviour in sinusoidal form is also seen in the amplitude variation. Figure 10 is a plot of the phase of light minimum against the observing year. Apart from the amplitude, the minima phases ($\theta_{min}$) also exhibit a variation. The minima phases are separating in two mean longitudes. The $\theta_{min}$ determined from such light curves would give the effective longitude of the spot or spots group. There is an indication of two effective longitudes in which the spots are generally formed. One of them is around 0$^{P}$.80, while the other is around 0$^{P}$.20.

\subsection{Flare Activity}

Apart from both eclipses and the sinusoidal-like variations out-of-eclipses, DK CVn also exhibits the flare activity. In this study, we detected one flare in the observations of 2007 and two flares in 2008. The flare shown in Figure 11 was detected in 2007 and exhibits itself in each band, while the flares detected in 2008 exhibits themselves in just B and V bands. They are shown in Figure 12. These two flares can not be detected in R band due to their lower powers. In addition to the flares detected in this study, the flare detected in B band by \citet{Ter05} was scanned and we got its observational data as well. Terrell's flare is shown in Figure 13. The flare parameters were calculated for each flare, and we list them in Table 5. In the table, the observing season, observing band, HJD of the flare maxima, flare rise time (s), flare decay time (s), flare total duration (s), flare equivalent duration (s), flare amplitude (mag) are listed in each column, respectively. In the last column, we noted which study the flare data are belonged.

To calculate the flare parameters, we used the method described by \citet{Dal10}. However, a different way was followed to determine the quiescent level of the brightness due to the eclipsing binary nature of DK CVn. Using the synthetic light curve; we obtained the quiescent level of the brightness for each phase. In order to test whether the method is correctly working or not, we compared the light curves observed in the consecutive-close nights and the synthetic light curve. Some examples for these comparisons are shown in Figures 11 and 12. As seen from the figures, DK CVn was observed two or more times in each phase intervals in close dates. Sometimes a flare was detected in one observing night, while no flare was detected in the same phase interval in another-close observing night. Comparing these observations with the synthetic light curve aids to determine the actual flare light variation.

As seen from Figure 11, a flare was detected around the primary minima on April 18, 2007. The system was observed in the same phase interval on March 8, 2007, but no flare was detected. The flare detected on April 18, 2007 distorted almost the shape of the primary minimum. The similar case is seen in the observations of 2008. Two flares were detected on March 16, 2008, while no flare was detected on March 2, 2008. The flare taken from \citet{Ter05} is shown in Figure 13. We compared Terrell's flare light curve with only the synthetic light curve due to absence of another observation in a consecutive-close night.

Using the Equations (2) and (3) described in the method developed by \citet{Ger72}, the flare equivalent durations and flare energies can be calculated. In Equation (2), $I_{0}$ is the intensity of the star in the quiescent level and $I_{flare}$ is the intensity during flare.

\begin{center}
\begin{equation}
P = \int[(I_{flare}-I_{0})/I_{0}] dt
\end{equation}
\end{center}

\begin{center}
\begin{equation}
E = P \times L
\end{equation}
\end{center}
where $E$ is the flare energy, $P$ is the flare equivalent duration given by Equation (2), and $L$ is the luminosity of the stars in the quiescent level.

To understand whether the flares observed from DK CVn, which is an eclipsing binary system, are similar to the flares occurring on the surface of UV Ceti type stars, DK CVn's flares were compared with B band flares of five UV Ceti type stars presented by \citet{Dal11}. To be able to compare them, first of all, following the method developed by \citet{Dal11}, the distribution of the flare equivalent durations versus flare total durations were derived for B band flares of DK CVn. Using SPSS V17.0 software \citep{Gre99} and GrahpPad Prism V5.02 software \citep{Mot07, Daw04}, the best model function was determined. Using the least-squares method, regression calculations showed that the best model function of distribution is the OPEA function \citep{Mot07, Spa87} given by Equation (4):

\begin{center}
\begin{equation}
y~=~y_{0}~+~(Plateau~-~y_{0})~\times~(1~-~e^{-k~\times~x})
\end{equation}
\end{center}

The derived OPEA model of DK CVn's flares is shown in Figure 14. Using this model, some parameters of the flare equivalent duration, such as $y_{0}$, $Plateau$, $K$, $Span$, $Half-Life$ values, were computed. In the OPEA model function given by Equation (4), the $y$ values were taken as flare equivalent duration in logarithmic scale, while $x$ values were taken as flare total durations. The parameter $y_{0}$ is the lowest flare equivalent duration obtained in logarithmic scale, while the parameter $Plateau$ is the upper limit the flare equivalent durations can reach. According to Equation (3), $Plateau$ value depends only on flare energy, while $y_{0}$ value depends on the brightness of the target and sensitivity of the optical system, as well as flare power. The parameter $K$ is a constant value depending on $x$ values. The $Span$ value is a difference between the $Plateau$ and $y_{0}$ values. The $Half-Life$ value is half of the first $x$ values, where the model reaches the $Plateau$ value. In other words, it is half of the flare total duration, where flares with the highest energy start to be seen. The $Half-Life$ value is an indicator of the duration the flare process occurring on the surface of a star needs to reach the saturation. As seen from the distribution of flare equivalent durations, the flare equivalent durations increase with the flare total duration until a specific total duration value, and then the flare equivalent durations became constant, no matter how long the flare total duration is. The $Half-Life$ value is half of this specific total duration value. All the parameters computed from the OPEA model are listed in Table 6. In order to test whether the $Plateau$ value is statistically acceptable for this distribution of the flare equivalent durations, using the Independent Sample t-Test (hereafter t-Test, \citealt{Mot07, Daw04, Gre99}), we compute the mean average value of the flares, which are located in the $Plateau$ phase of the model. The found mean average value is listed in the last row of Table 6.

All the parameters listed in Table 6 were compared with the parameters given by \citet{Dal11} for five UV Ceti type stars. In the comparison, we assume that the flares are occurring on the surface of the cool component of the system. The light curve analysis of the system indicates that the temperature of the cool component is 3123 K. According to \citet{Tok00}, this temperature is corresponding to B-V = $1^{m}.630$ as the colour index. All the comparisons are shown in Figures 15 and 16.

As seen from Figures 15 and 16, in fact, the flares detected in the observations of DK CVn seem to have a same nature with the flares detected from UV Ceti type stars. Moreover, our assumption also seems to be correct, because the parameters are in agreement with its analogue according to assumed temperature.

\section{Results and Discussion}

In this study, we obtained the light curves of an eclipsing binary system, DK CVn, in two observing seasons, and we analysed the BVR light curves obtained in 2008 to find the physical properties of DK CVn. In addition, using Kepler's third law under some assumptions, the possible absolute parameters were found. The observations demonstrate that the radii are generally larger than the expected values. We compared the radii of DK CVn's components with the known active stars and a model. The radii of the active stars, which exhibit spot or flare activity or both of them, are dramatically larger than the values given by the model developed for the stars with $Z=0.02$ by \citet{Sie00}. The components of DK CVn are also in agreement with the other active stars, listed in the catalogue of \citet{Ger99}. According to several theoretical models and observational studies \citep{Rib06, Cha07, Mor08, Mor10}, the case seen in Figure 7 is a well known phenomenon for low-mass active stars. For instance, YY Gem \citep{Tor02}, CU Cam \citep{Rib03} and CM Dra \citep{Mor09} are the most popular system for this case. There are several similarities between these three systems and DK CVn. Firstly, all of them exhibit the spot and flare activities, and they consist of the low mass components.

The observations in B and V bands demonstrate that the system exhibits flare activity. Considering the effective temperature, we assumed that the flares occur on the secondary component. The derived parameters demonstrate that DK CVn's flares seem to behave in the same way with the flares of UV Ceti stars. The maximum energy level of the flares seen in the DK CVn system is in agreement with the analogues of UV Ceti stars from the late spectral types. Consequently, the flares detected from DK Cvn must be produced by the same process occurring on the surface of an UV Ceti star.

Considering the flare activity exhibited by DK CVn system, both components could be chromospherically active stars. In addition, the spectral types of both components are also supported to the chromospherical activity. Another support comes from the light curve analysis. The observed light curve can be modelled with two cool spots on the one of the components due to the asymmetry seen in the shape of the curve. We assumed that the spots are located on the primary component, and we analysed the light curves with this assumption. According to its effective temperature, the secondary component is so close the border of the full-convective area among the M dwarfs. Although the full-convective M dwarfs exhibit very strong flare activity, a few of them just exhibit spot activity. However, the K dwarfs are generally potential stars, which are possible to exhibit spot activity. This is why we assumed the spotted star is the primary component.

Considering the sum of weighted squared residuals, we found from the BVR light curve analyses that there are a large spot and a small one on the primary component. The larger spot with a temperature factor of 0.95 is located in longitude of 188$^\circ$, while the small one with a temperature factor of 0.90 is located in longitude of 290$^\circ$. It is well known that the longitudes of the spotted areas can be found exactly from the light curves obtained with photometric observations. However, the co-latitude, radius and temperature factor of a spot are not very well. A similar synthetic curves fitting the observations can be derived with some different values of these parameters. In this study, when we considered the sum of weighted squared residuals, the radii, co-latitudes and temperature factors derived for the spots give one of the best synthetic curves, which are seen in Figure 5. They are statistically acceptable, and the physical parameters of the components are also acceptable in the astrophysical sense with these spot parameters.

However, the asymmetry and variation on the pre-whitened light curves reveal that cool spots vary with time, as well. Although the spots can be sometimes changing in few months as it is seen from NSVS data, generally the spots do not seem to change on the short time-scale. The spots occur at two longitudes, i.e., in phases of 0.80-0.90 and 0.00-0.20. The phenomenon reveals that spots concentrate on two active longitudes. The amplitude of the pre-whitened light curves demonstrated very well variation. In general, the amplitudes of the variations at out-of-eclipses have been decreasing since 2002, while a cyclic variation in sinusoidal shape is also seen, combining with the general decreasing.

The phases of the minima are dramatically changing in this system. The variations of the pre-whitened light curves indicate that the spotted areas are not stable on the component. In the pre-whitened light curves of the season 2003, two minima are seen separately from each other. In the season 2005, the pre-whitened light curve has a very strong asymmetry. The variations seen out-of-eclipses are similar to the variations exhibiting by the young-fast-rotating stars, such as YY Gem, ER Vul, SV Cam, CU Cam and CM Dra \citep{Str09, Tor02, Rib03, Mor09}. Therefore, the spotted component (and the system) could be a young star. On the other hand, it must be noted that there are many systems exhibit an unexpected cases in contrast to this approach \citep{Roc02}.

In brief, the variation seen at out-of-eclipses should be due to the cool spots gathering into two separated longitudes on the surface of the primary component. Thus, this variation is caused by the rotational modulation due to the chromospherical activity. Consequently, DK CVn seems to be an analogue of RS CVn type stars. Besides, both spot and flare activities indicate that the system has high level chromospherical activity.

\section*{Acknowledgments} The authors acknowledge generous allotments of observing time at the Ege University Observatory. We thank Dr. Dirk Terrell for reviewing the manuscript and his useful comments that have contributed to the improvement of the paper.

\clearpage

\begin{figure*}[h]
\hspace{0.8cm}
\includegraphics[scale=0.85, angle=0]{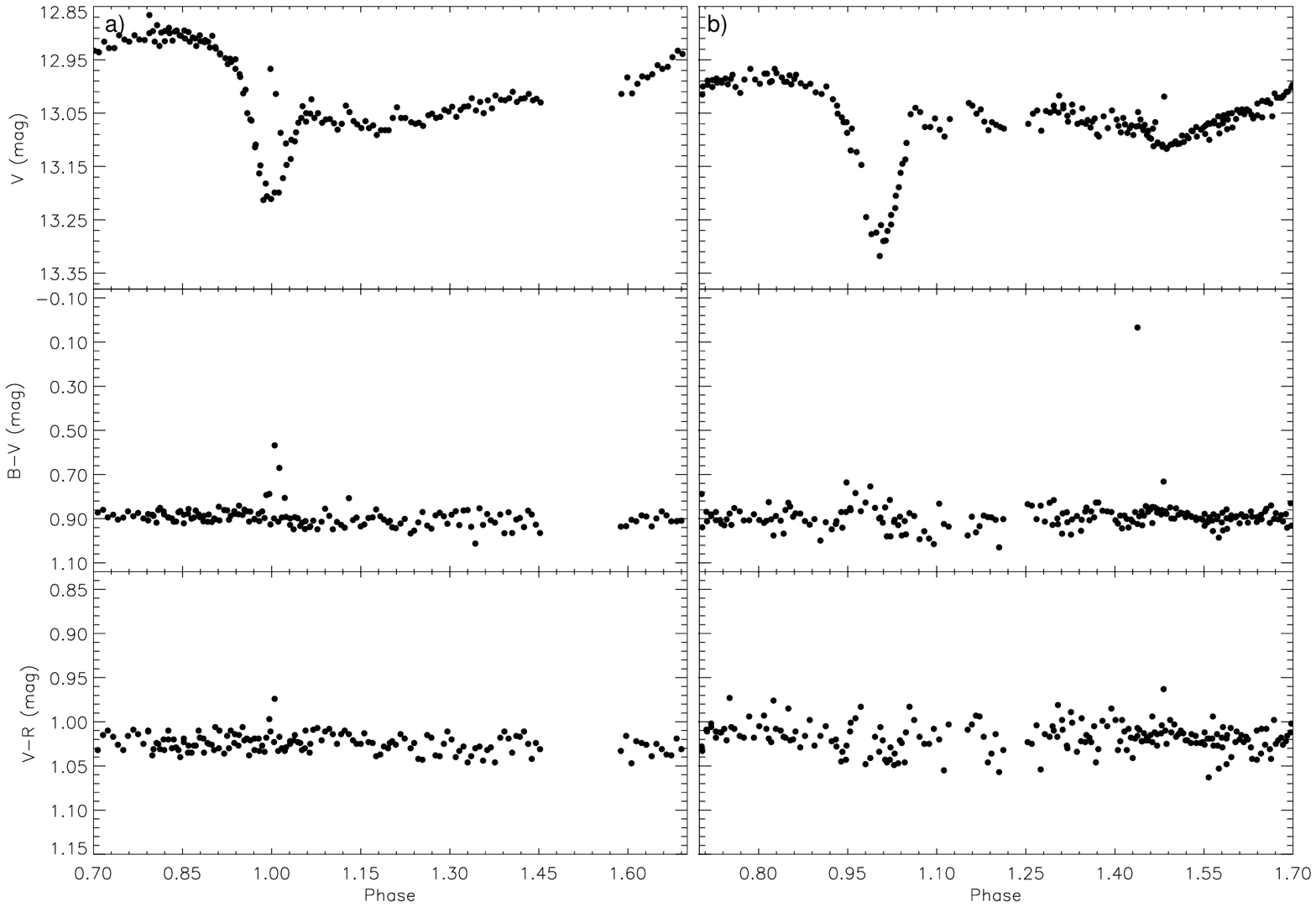}
\vspace{0.3cm}
\caption{The V band light curve with both B-V and V-R colour curves of DK CVn. a) The observations of the season 2007. b) The observations of the season 2008.}
\label{Fig1}
\end{figure*}

\begin{figure*}[h]
\hspace{2.6cm}
\includegraphics[scale=0.90, angle=0]{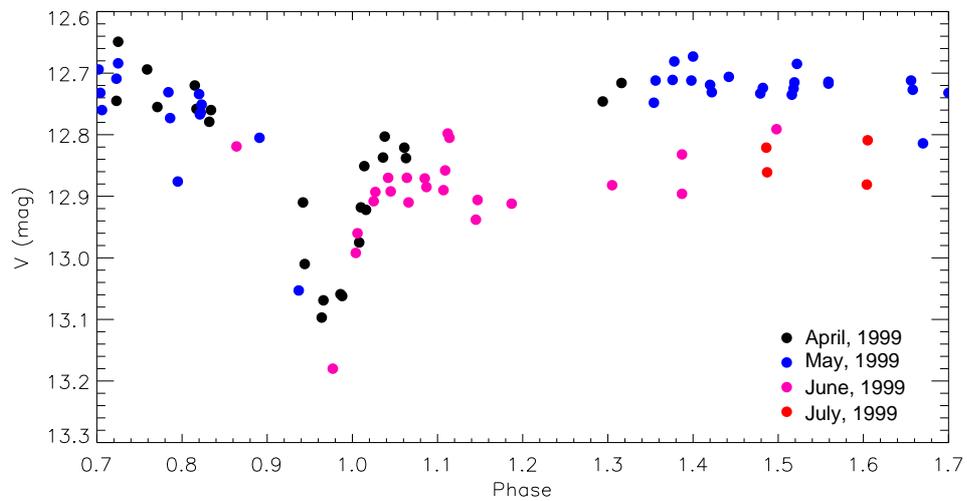}
\vspace{0.3cm}
\caption{The V band light curves of the observing season 1999. The data were taken from the NSVS database \citep{Woz04}.}
\label{Fig2}
\end{figure*}

\begin{figure*}[h]
\hspace{3cm}
\includegraphics[scale=0.90, angle=0]{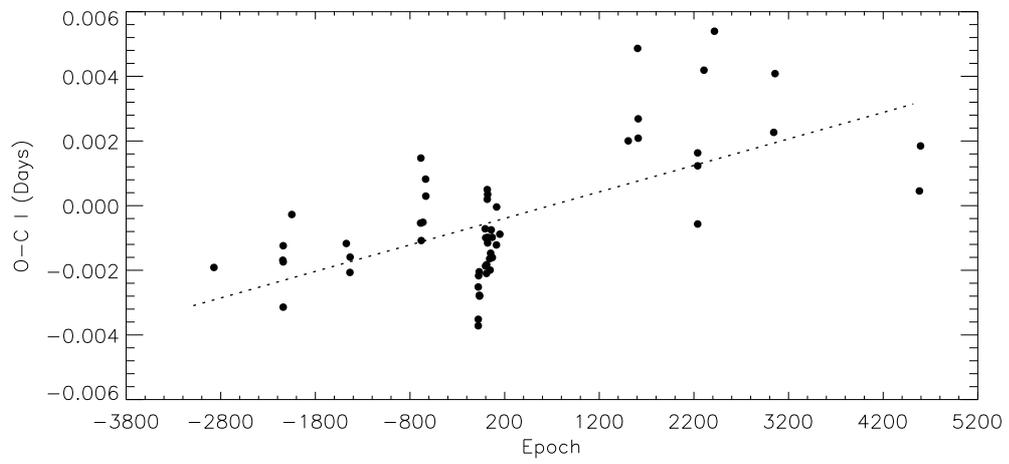}
\vspace{0.3cm}
\caption{DK CVn's $O-C$ diagram (The dashed line represents the linear fit).}
\label{Fig3}
\end{figure*}

\begin{figure*}[h]
\hspace{3.8cm}
\includegraphics[scale=0.88, angle=0]{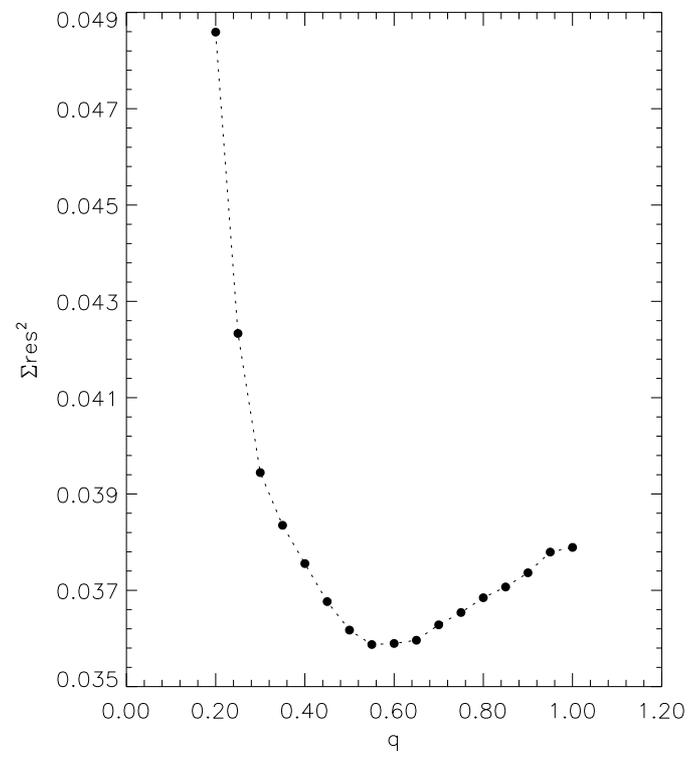}
\vspace{0.3cm}
\caption{The variation of the sum of weighted squared residuals versus mass ratio in the "q search".}
\label{Fig4}
\end{figure*}

\begin{figure*}[h]
\hspace{4.2cm}
\includegraphics[scale=0.90, angle=0]{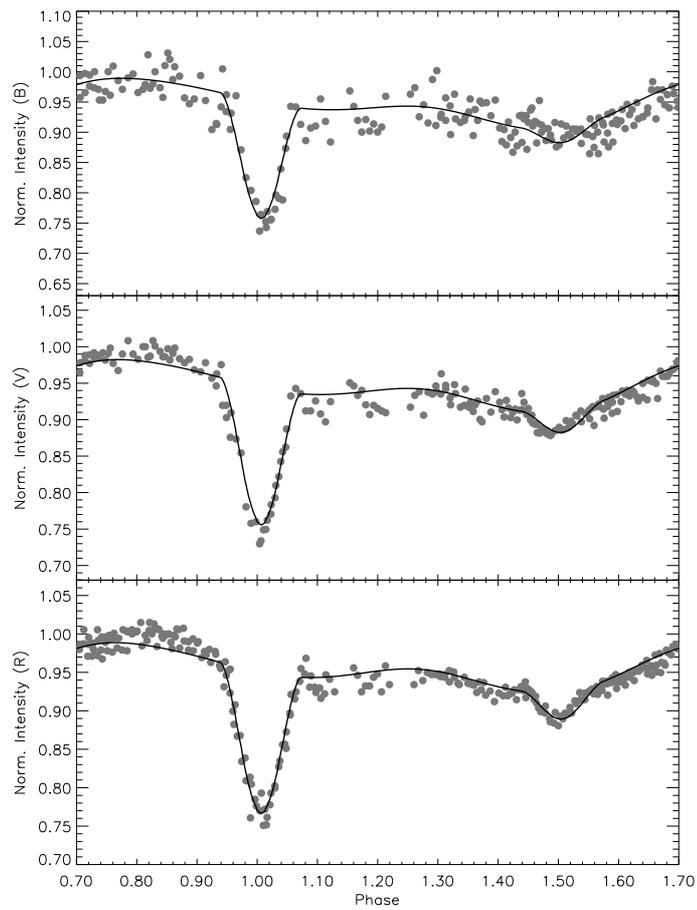}
\vspace{0.3cm}
\caption{DK CVn's BVR band light curves (filled circles) observed in 2008 and the synthetic curves (lines) derived from the light curve solution.}
\label{Fig5}
\end{figure*}

\begin{figure*}[h]
\hspace{2.6 cm}
\includegraphics[scale=0.50, angle=0]{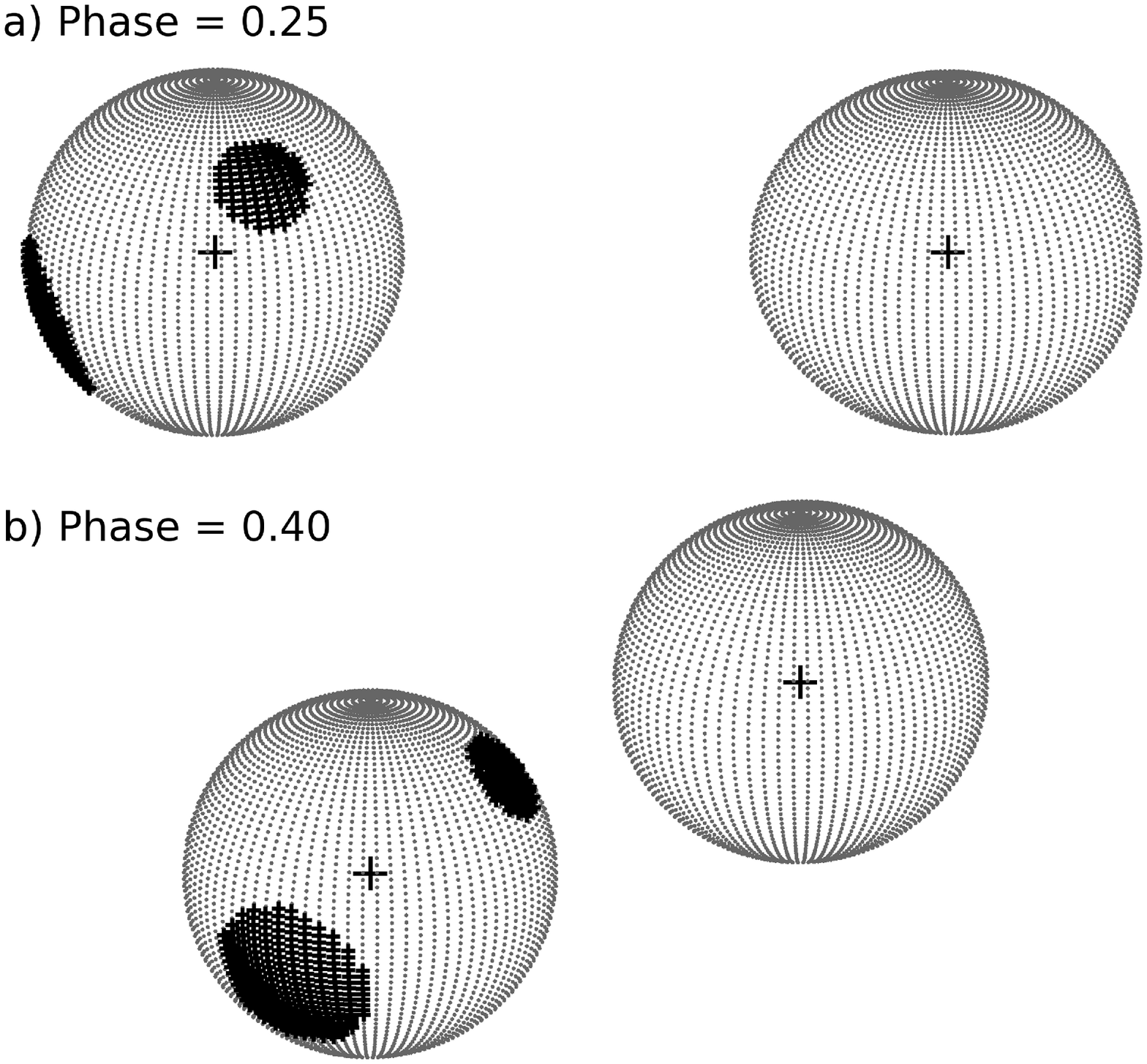}
\caption{The geometric configurations at the (a) phase 0.25 and (b) 0.40, illustrated for DK CVn.}
\label{Fig6}
\end{figure*}

\begin{figure*}[h]
\hspace{4.8cm}
\includegraphics[scale=0.90, angle=0]{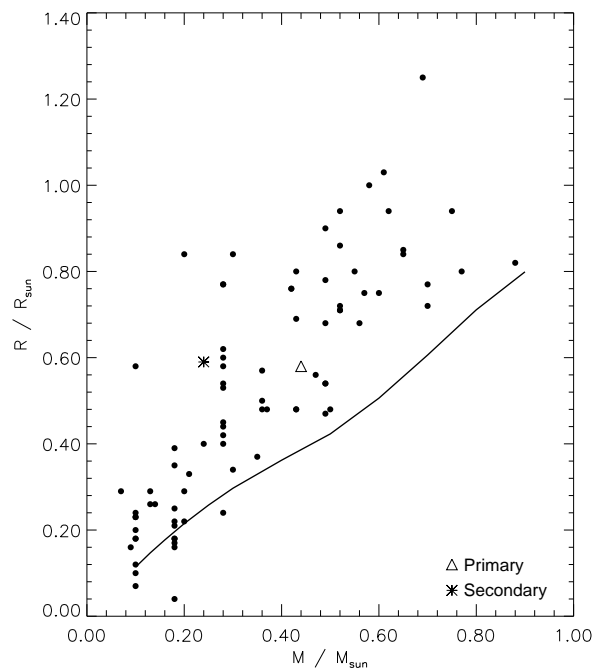}
\vspace{0.3cm}
\caption{The places of the components of DK CVn among UV Ceti type stars in the Mass-Radius distribution. In the figure, the filled circles represent the active stars listed in the catalogue of \citet{Ger99}. The asterisk represents the secondary component, while the open triangle represents the primary component of DK CVn. The line represents the ZAMS theoretical model developed by \citet{Sie00}.}
\label{Fig7}
\end{figure*}

\begin{figure*}[h]
\hspace{1cm}
\includegraphics[scale=0.85, angle=0]{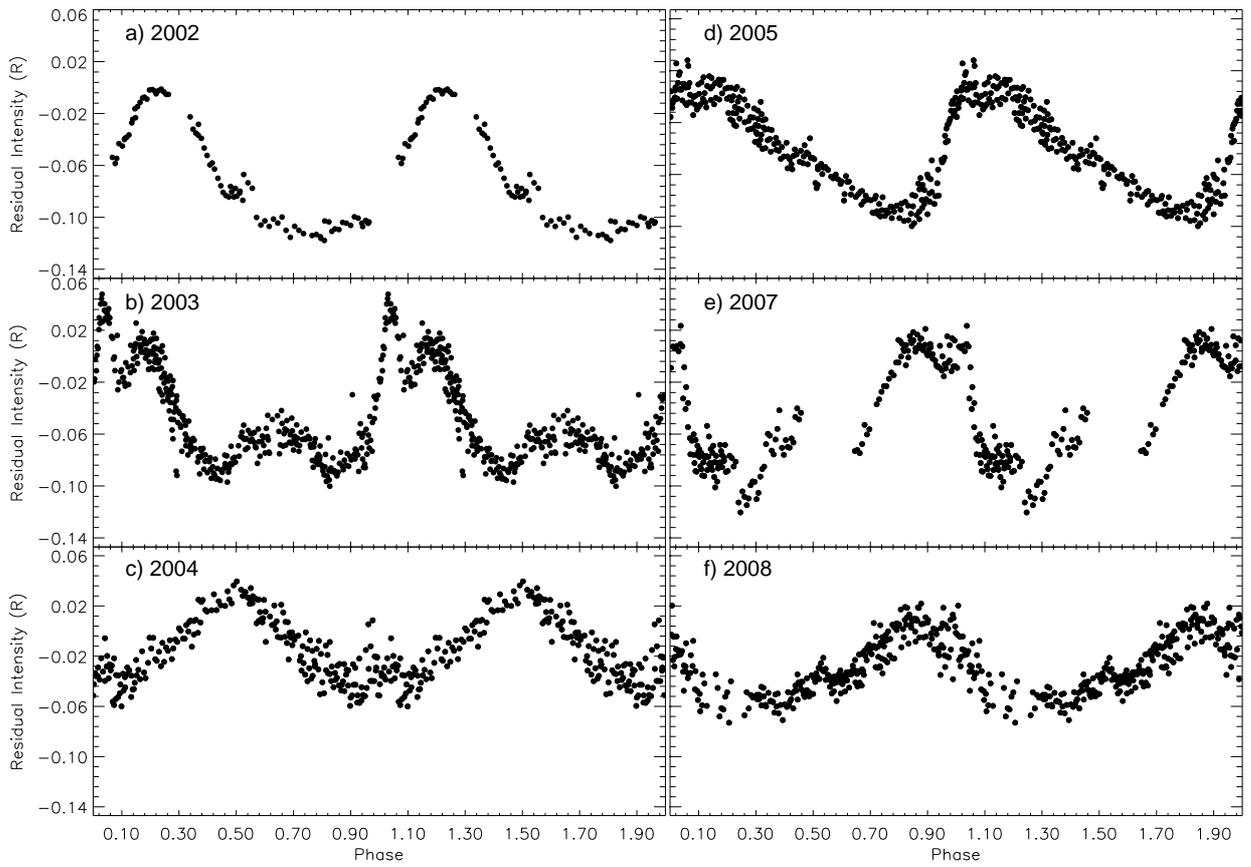}
\vspace{0.3cm}
\caption{All the pre-whitened light curves in R band. (All the light curves are shown as double cycle for better visibility of light variations).}
\label{Fig8}
\end{figure*}

\begin{figure*}[h]
\hspace{2.6cm}
\includegraphics[scale=0.90, angle=0]{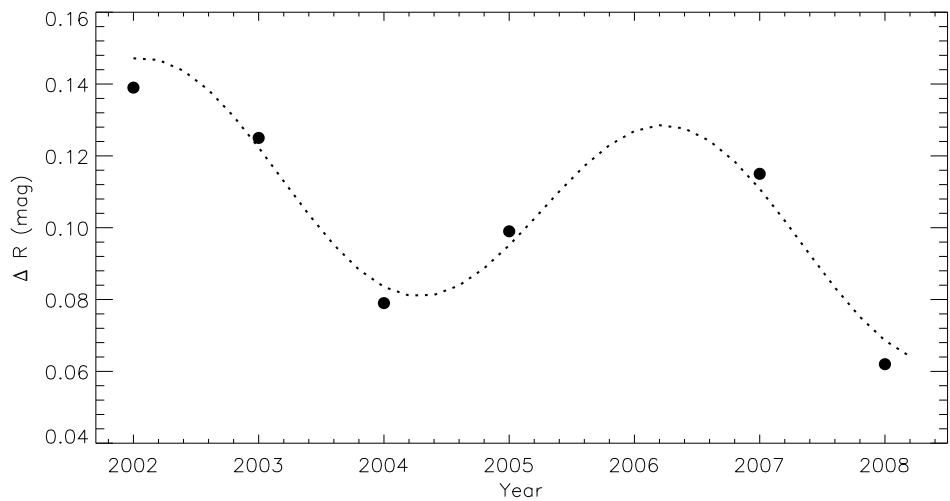}
\vspace{0.3cm}
\caption{The variation of the amplitude of the variation at out-of-eclipses throughout the years. The dashed line represents the fits of the variations in the figure.}
\label{Fig9}
\end{figure*}

\begin{figure*}[h]
\hspace{2.2cm}
\includegraphics[scale=1.00, angle=0]{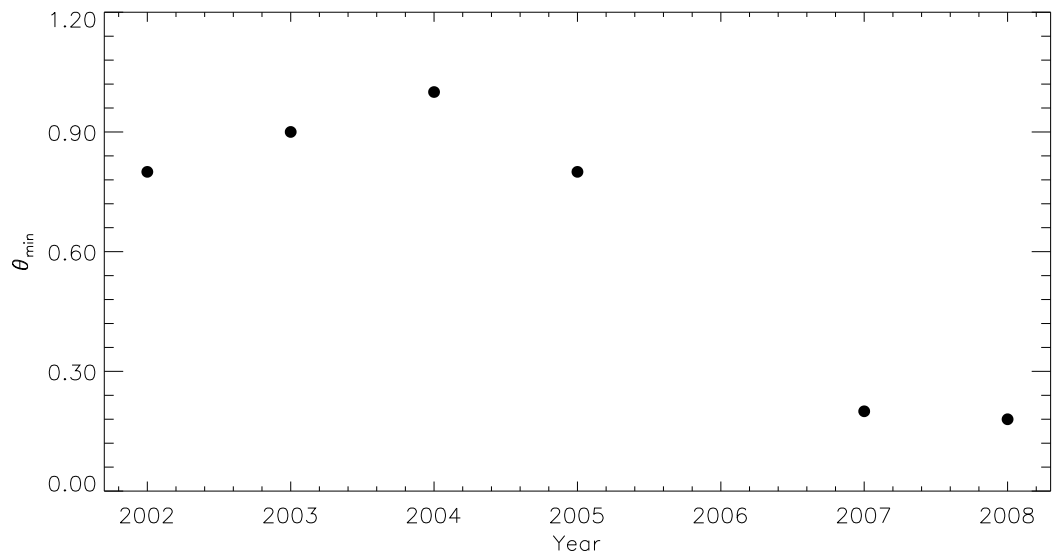}
\vspace{0.3cm}
\caption{The minimum phases for the variation at out-of-eclipses throughout the years.}
\label{Fig10}
\end{figure*}

\begin{figure*}[h]
\hspace{4.4cm}
\includegraphics[scale=0.90, angle=0]{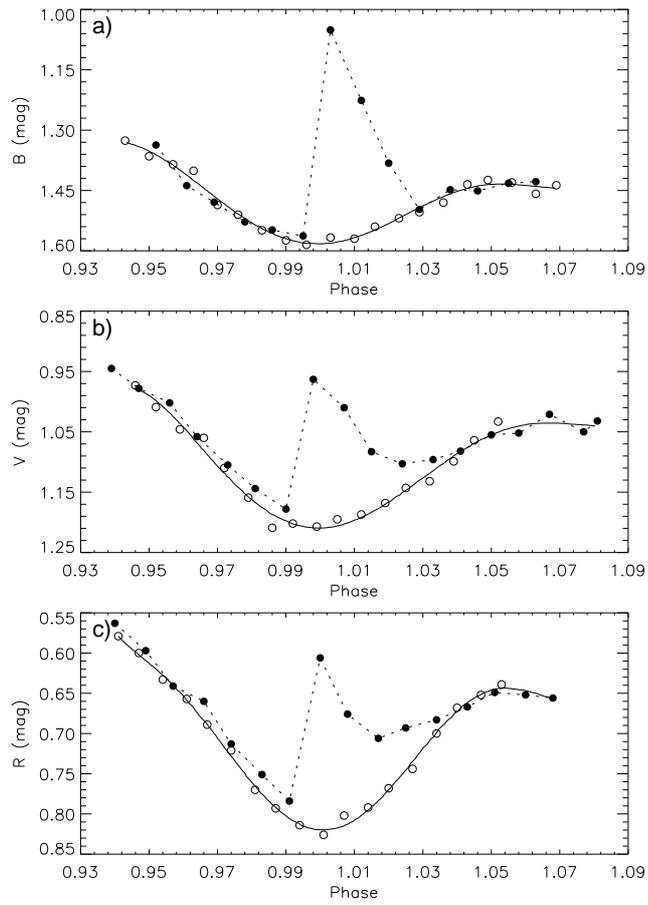}
\vspace{0.3cm}
\caption{The fast flare sample detected around primary minimum on April 18, 2007. In figure, open circles represent the observations on April 18, 2007, while filled circles represent the observations on March 8, 2007. The line represents the synthetic light curve.}
\label{Fig11}
\end{figure*}

\begin{figure*}[h]
\hspace{4.4cm}
\includegraphics[scale=0.90, angle=0]{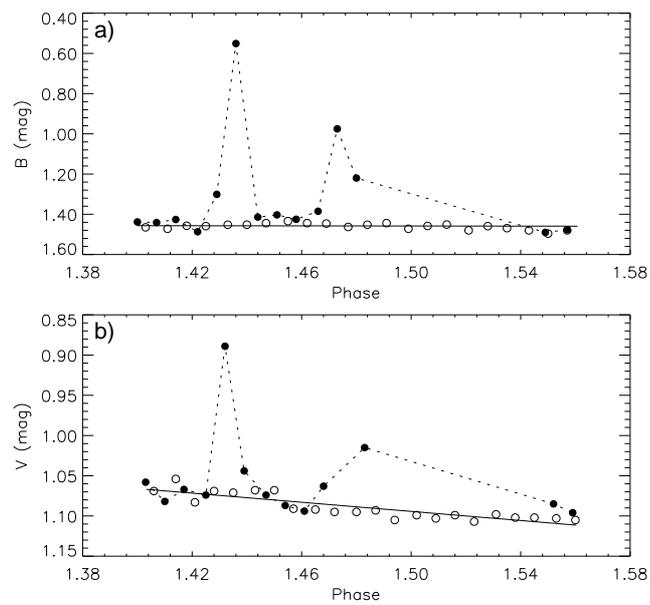}
\vspace{0.3cm}
\caption{Two fast flare samples detected on March 16, 2008. In figure, open circles represent the observations on March 16, 2008, while filled circles represent the observations on March 2, 2008. The line represents the synthetic light curve.}
\label{Fig12}
\end{figure*}

\begin{figure*}[h]
\hspace{4.3cm}
\includegraphics[scale=1.00, angle=0]{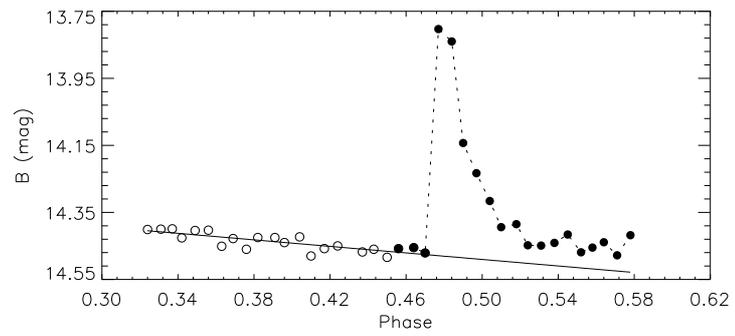}
\vspace{0.3cm}
\caption{The fast flare sample detected by \citet{Ter05} in B band.}
\label{Fig13}
\end{figure*}

\begin{figure*}[h]
\hspace{3.2cm}
\includegraphics[scale=0.80, angle=0]{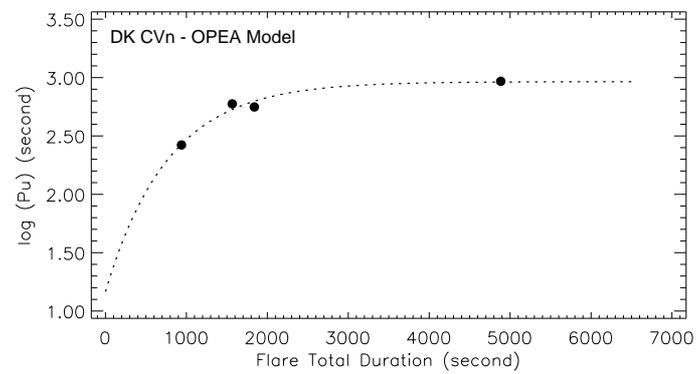}
\vspace{0.3cm}
\caption{The distribution of the flare equivalent durations versus the flare total durations for DK CVn flares (filled circles). The OPEA model (dashed line) derived for this distribution.}
\label{Fig14}
\end{figure*}

\begin{figure*}[h]
\hspace{0.84cm}
\includegraphics[scale=0.85, angle=0]{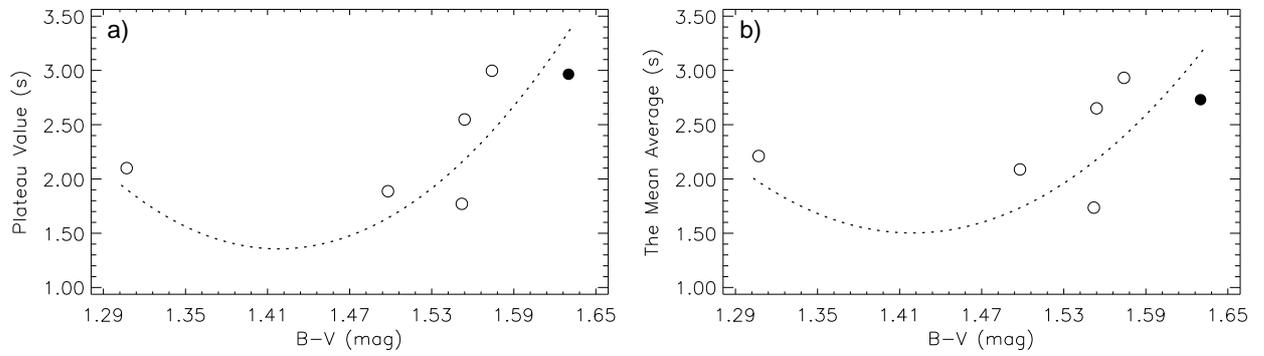}
\vspace{0.3cm}
\caption{The variations of (a) the $Plateau$ values derived from the OPEA models and (b) the mean equivalent durations computed by t-Test analyses versus B-V colour index. In the figures, the filled circles represent the DK CVn, while open circles represent five UV Ceti type stars taken from \citet{Dal11}. The dashed lines are just used to show the variation trend.}
\label{Fig15}
\end{figure*}

\begin{figure*}[h]
\hspace{0.84cm}
\includegraphics[scale=0.85, angle=0]{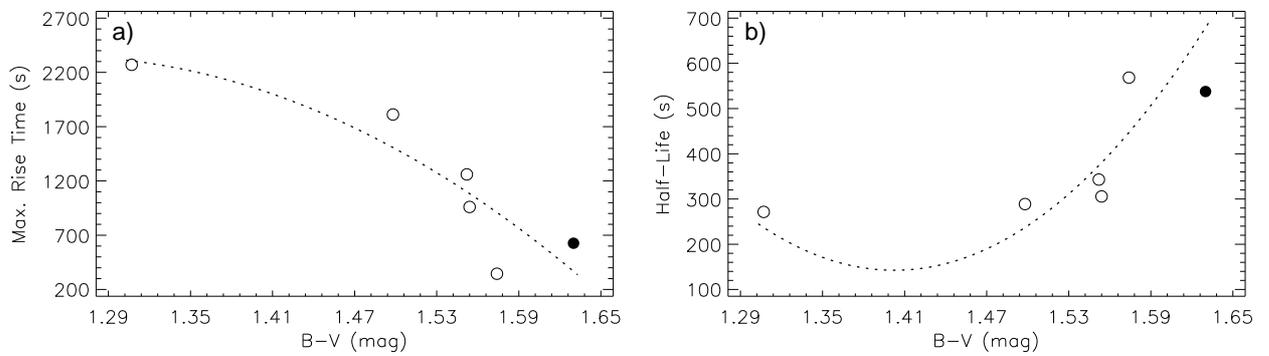}
\vspace{0.3cm}
\caption{The variations of (a) the maximum flare rise times and (b) the $Half-Life$ values versus B-V colour index. In the figures, all the symbols are the same with Figure 15.}
\label{Fig16}
\end{figure*}

\clearpage

\begin{table*}
\begin{center}
\caption{Basic parameters for the observed stars. V band brightness and B-V indexes were obtained in this study. Considering B-V indexes, the spectral types were taken from \citet{Tok00}.\label{tbl-1}}
\vspace{0.3 cm}
\begin{tabular}{@{}lcccc@{}}
\hline\hline
Star	&	Alpha / Delta (J2000)	&	V (mag)	&	B-V (mag)	&	Spectral Type	\\
\hline									
DK CVn	&	12$^{h}$ 33$^{m}$ 09$^{s}$.34 / +37$^{\circ}$ 58$^{\prime}$ 20$^{\prime\prime}$.28	&	12.967	&	0.890	&	K2	\\
GSC 3018 2499	&	12$^{h}$ 33$^{m}$ 11$^{s}$.14 / +37$^{\circ}$ 45$^{\prime}$ 12$^{\prime\prime}$.90	&	12.004	&	0.527	&	F8	\\
GSC 3018 2425	&	12$^{h}$ 32$^{m}$ 58$^{s}$.36 / +37$^{\circ}$ 54$^{\prime}$ 20$^{\prime\prime}$.30	&	12.702	&	0.628	&	G2	\\
\hline
\end{tabular}
\end{center}
\end{table*}

\begin{table*}															
\begin{center}															
\caption{The minima times and $O-C$ residuals.\label{tbl-2}}														
\vspace{0.3 cm}															
\begin{tabular}{@{}ccccccc@{}}															
\hline\hline															
HJD (+24 00000)	&	E	&	$(O-C)_{I}$	&	$(O-C)_{II}$	&	Filter	&	Ref.	\\
\hline											
52001.4487	&	-2872.0	&	-0.0019	&	0.0010	&	CCD	&	1	\\
52361.7820	&	-2144.0	&	-0.0017	&	0.0006	&	R	&	2	\\
52363.7604	&	-2140.0	&	-0.0031	&	-0.0008	&	I	&	2	\\
52363.7618	&	-2140.0	&	-0.0017	&	0.0006	&	R	&	2	\\
52363.7623	&	-2140.0	&	-0.0012	&	0.0011	&	V	&	2	\\
52408.8049	&	-2049.0	&	-0.0003	&	0.0020	&	U	&	2	\\
52693.9027	&	-1473.0	&	-0.0012	&	0.0006	&	CCD	&	2	\\
52712.7104	&	-1435.0	&	-0.0021	&	-0.0003	&	R	&	2	\\
52713.7008	&	-1433.0	&	-0.0016	&	0.0002	&	R	&	2	\\
53082.4493	&	-688.0	&	-0.0005	&	0.0006	&	CCD	&	2	\\
53083.9362	&	-685.0	&	0.0015	&	0.0026	&	R	&	2	\\
53085.9135	&	-681.0	&	-0.0011	&	0.0000	&	V	&	2	\\
53094.8234	&	-663.0	&	-0.0005	&	0.0006	&	V	&	2	\\
53108.6837	&	-635.0	&	0.0008	&	0.0019	&	I	&	2	\\
53109.6731	&	-633.0	&	0.0003	&	0.0014	&	B	&	2	\\
53383.8786	&	-79.0	&	-0.0037	&	-0.0031	&	V	&	2	\\
53383.8788	&	-79.0	&	-0.0035	&	-0.0029	&	V	&	2	\\
53383.8798	&	-79.0	&	-0.0025	&	-0.0019	&	R	&	2	\\
53385.8600	&	-75.0	&	-0.0022	&	-0.0015	&	V	&	2	\\
53388.8299	&	-69.0	&	-0.0020	&	-0.0014	&	R	&	2	\\
53389.8191	&	-67.0	&	-0.0028	&	-0.0022	&	R	&	2	\\
53390.8090	&	-65.0	&	-0.0028	&	-0.0022	&	R	&	2	\\
53420.0139	&	-6.0	&	-0.0007	&	-0.0001	&	B	&	2	\\
53421.9926	&	-2.0	&	-0.0019	&	-0.0013	&	B	&	2	\\
53422.9834	&	0.0	&	-0.0010	&	-0.0004	&	B	&	2	\\
53426.9420	&	8.0	&	-0.0021	&	-0.0015	&	B	&	2	\\
53427.9322	&	10.0	&	-0.0018	&	-0.0013	&	B	&	2	\\
53430.9040	&	16.0	&	0.0002	&	0.0007	&	V	&	2	\\
53430.9043	&	16.0	&	0.0005	&	0.0010	&	B	&	2	\\
53432.8825	&	20.0	&	-0.0012	&	-0.0006	&	B	&	2	\\
53432.8840	&	20.0	&	0.0003	&	0.0009	&	V	&	2	\\
53433.8726	&	22.0	&	-0.0010	&	-0.0004	&	B	&	2	\\
53443.7712	&	42.0	&	-0.0016	&	-0.0011	&	V	&	2	\\
53445.7507	&	46.0	&	-0.0020	&	-0.0015	&	B	&	2	\\
53448.7210	&	52.0	&	-0.0015	&	-0.0010	&	B	&	2	\\
53451.6915	&	58.0	&	-0.0007	&	-0.0002	&	V	&	2	\\
53456.6409	&	68.0	&	-0.0010	&	-0.0005	&	R	&	2	\\
53457.6302	&	70.0	&	-0.0016	&	-0.0011	&	B	&	2	\\
53478.9140	&	113.0	&	-0.0012	&	-0.0007	&	R	&	2	\\
53479.9051	&	115.0	&	0.0000	&	0.0004	&	B	&	2	\\
53496.7330	&	149.0	&	-0.0009	&	-0.0004	&	R	&	2	\\
54168.4007	&	1506.0	&	0.0020	&	0.0013	&	BVR	&	3	\\
54217.4049	&	1605.0	&	0.0049	&	0.0041	&	VR	&	3	\\
54220.3719	&	1611.0	&	0.0021	&	0.0013	&	R	&	4	\\
54220.3725	&	1611.0	&	0.0027	&	0.0019	&	VR	&	3	\\
54531.4535	&	2239.5	&	-0.0006	&	-0.0019	&	V	&	5	\\
54531.4553	&	2239.5	&	0.0012	&	-0.0001	&	I	&	5	\\
54531.4557	&	2239.5	&	0.0016	&	0.0003	&	R	&	5	\\
54564.3733	&	2306.0	&	0.0042	&	0.0029	&	BVR	&	3	\\
54619.3154	&	2417.0	&	0.0054	&	0.0040	&	VR	&	3	\\
54929.4066	&	3043.5	&	0.0023	&	0.0003	&	R	&	4	\\
54936.3379	&	3057.5	&	0.0041	&	0.0021	&	CCD	&	4	\\
55691.4004	&	4583.0	&	0.0005	&	-0.0028	&	VR	&	6	\\
55697.3413	&	4595.0	&	0.0018	&	-0.0014	&	RI	&	6	\\
\hline													
\end{tabular}													
\end{center}				
$^{1}$	\citet{Bra07}			\\
$^{2}$	\citet{Ter05}			\\
$^{3}$	\citet{Sip09}			\\
$^{4}$	\citet{Bra09}			\\
$^{5}$	\citet{Bra08}			\\
$^{6}$	This Study			\\
\end{table*}

\begin{table*}
\begin{center}
\caption{The parameters obtained from the light curve analysis.\label{tbl-3}}
\vspace{0.3 cm}
\begin{tabular}{@{}lc@{}}
\hline\hline
Parameter	&	Value	\\
\hline							
$q$	&	0.55 (Fixed)	\\
$i$ ($^\circ$)	&	71.12$\pm$0.04	\\
$T_{1}$ (K)	&	4040	\\
$T_{2}$ (K) 	&	3123$\pm$16	\\
$\Omega_{1}$	&	4.542$\pm$0.003	\\
$\Omega_{2}$	&	3.424$\pm$0.001	\\
L$_{1}$/L$_{T}$ $(B)$	&	0.848$\pm$0.004	\\
L$_{1}$/L$_{T}$ $(V)$	&	0.909$\pm$0.002	\\
L$_{1}$/L$_{T}$ $(R)$	&	0.883$\pm$0.003	\\
$g_{1}$	&	0.32 (Fixed)	\\
$g_{2}$	&	0.32 (Fixed)	\\
$A_{1}$	&	0.50 (Fixed)	\\
$A_{2}$	&	0.50 (Fixed)	\\
$x_{1,bol}$	&	0.563 (Fixed) \\
$x_{1,B}$	&	0.826 (Fixed) \\
$x_{1,V}$	&	0.799 (Fixed) \\
$x_{1,R}$	&	0.747 (Fixed) \\
$x_{2,bol}$	&	0.468 (Fixed) \\
$x_{2,B}$	&	0.868 (Fixed) \\
$x_{2,V}$	&	0.839 (Fixed) \\
$x_{2,R}$	&	0.748 (Fixed) \\
$<r_{1}>$	&	0.254$\pm$0.005	\\
$<r_{2}>$	&	0.258$\pm$0.004	\\
$Co-Lat_{Spot~I}$ ($^{\circ}$) 	&	 110 (Fixed)	\\
$Long_{Spot~I}$ ($^{\circ}$) 	&	 188 (Fixed)	\\
$R_{Spot~I}$ ($^{\circ}$) 	&	 26 (Fixed)	\\
$T_{eff, Spot~I}$ 	&	 0.95 (Fixed)	\\
$Co-Lat_{Spot~II}$ ($^{\circ}$) 	&	 50 (Fixed)	\\
$Long_{Spot~II}$ ($^{\circ}$) 	&	 290 (Fixed)	\\
$R_{Spot~II}$ ($^{\circ}$) 	&	 15 (Fixed)	\\
$T_{eff,~Spot~II}$ 	&	 0.90 (Fixed)	\\
\hline
\end{tabular}
\end{center}
\end{table*}

\begin{table*}
\begin{center}
\caption{The minimum phases and the amplitudes of the sinusoidal-like variations seen in the pre-whitened light curves in R band.\label{tbl-4}}
\vspace{0.3 cm}
\begin{tabular}{@{}cccr@{}}
\hline\hline
Year	&	$\theta_{min}$	&	Amplitude in R (mag)	&	Ref	\\
\hline									
2002	&	0.80	&	0.139	&	1	\\
2003	&	0.90	&	0.125	&	1	\\
2004	&	1.00	&	0.079	&	1	\\
2005	&	0.80	&	0.099	&	1	\\
2007	&	0.20	&	0.115	&	2	\\
2008	&	0.18	&	0.062	&	2	\\
\hline
\end{tabular}
\end{center}
$^{1}$ \citet{Ter05} \\
$^{2}$ This Study \\
\end{table*}

\begin{table*}
\begin{center}
\caption{The flare parameters obtained from the flares detected in this study and the flare detected by \citet{Ter05}.\label{tbl-5}}
\begin{tabular}{@{}ccccccccc@{}}
\hline\hline
Year &	Filter &	HJD of Maxima &	 Rise	&	Decay	&	Total	&	Equivalent	&	Amplitude	&	Ref. \\
	&		&	(+24 00000)	&	Time (s)	&	Time (s)	&	Duration (s)	&	Duration (s)	&	(mag)	&	\\	
\hline															
2005	&	B	&	53394.90620	&	568	&	4315	&	4884	&	929.1551	&	0.668	&	1	\\
2007	&	B	&	54209.48337	&	363	&	1476	&	1839	&	429.6051	&	0.516	&	2	\\
2008	&	B	&	54542.31363	&	626	&	940	&	1566	&	595.2623	&	0.901	&	2	\\
2008	&	B	&	54542.33176	&	313	&	626	&	939	&	264.5768	&	0.488	&	2	\\
2007	&	V	&	54209.48084	&	363	&	1476	&	1839	&	214.9586	&	0.244	&	2	\\
2008	&	V	&	54542.31137	&	313	&	627	&	940	&	68.9973	&	0.180	&	2	\\
2007	&	R	&	54209.48168	&	363	&	1476	&	1839	&	189.9065	&	0.220	&	2	\\
\hline
\end{tabular}
\end{center}
$^{1}$ \citet{Ter05} \\
$^{2}$ This study. \\
\end{table*}

\begin{table*}
\begin{center}
\caption{The parameters computed from the OPEA model of DK CVn's flares.\label{tbl-6}}
\vspace{0.3 cm}
\begin{tabular}{@{}lc@{}}
\hline\hline
Parametre	&	Value	\\
\hline
Max. Rise Time (s)	&	626	\\
Max. Tot. Time (s)	&	4884	\\
$Plateau$ (s)	&	2.965 $\pm$ 0.07374	\\
$y_{0}$ (s)	&	1.167 $\pm$ 0.1555	\\
$K$ (s)	&	0.001289 $\pm$ 0.000495	\\
$Span$ (s)	&	1.798 $\pm$ 0.1243	\\
$Half - Life$ (s)	&	537.7	\\
Mean (s)	&	2.731	\\
\hline
\end{tabular}
\end{center}
\end{table*}


\end{document}